\documentclass[%
 superscriptaddress,
 preprintnumbers,
 nature,npj,
 amsmath,amssymb,
]{revtex4-1}

\usepackage{revquantum}
\usepackage{multirow}
\usepackage{hyperref}
\usepackage{academicons}
\usepackage{xcolor}

\newcommand{\orcid}[1]{\href{https://orcid.org/#1}{\textcolor[HTML]{A6CE39}{\aiOrcid}}}

\hypersetup{colorlinks,allcolors=blue}

\usepackage{graphicx}
\usepackage{subcaption} 
\usepackage{hyperref}
\usepackage{color}
\definecolor{darkblue}{rgb}{0,0,0.8}
\hypersetup{
	colorlinks=true,
	linkcolor=darkblue,
	citecolor=darkblue,
	urlcolor=darkblue
}
\usepackage{enumitem}
\usepackage{calc}
\usepackage{wrapfig}
\usepackage{bm}

\usepackage{color}
\definecolor{red}{rgb}{1,0, 0}
\definecolor{bzgcol}{rgb}{0.7,0, 0}
\definecolor{ascol}{rgb}{0,0.4, 0}
\definecolor{fkcol}{rgb}{0,0, 0.7}
\definecolor{incol}{rgb}{0.1,0.7, 0.7}

\newcommand{\angstrom}{\text{\normalfont\AA}}

\graphicspath{{other/}}

\newcommand{\mmtp}{{\rm mMTP}}
\newcommand{\mtp}{{\rm MTP}}
\newcommand{\dft}{{\rm DFT}}

\renewcommand{\_}{\char`_}

\usepackage[utf8]{inputenc}
\usepackage{xcolor}
\usepackage{graphicx}
\usepackage{dcolumn}

\usepackage{array}
\newcolumntype{P}[1]{>{\centering\arraybackslash}p{#1}}

\usepackage{subdepth}

\definecolor{rev0}{rgb}{0,0,0}

\begin{document}

\title{Magnetic Moment Tensor Potentials for collinear spin-polarized materials reproduce different magnetic states of bcc Fe}

\author{Ivan Novikov *}
\affiliation{Skolkovo Institute of Science and Technology, Skolkovo Innovation Center, Nobel St. 3, Moscow 143026, Russia}
\affiliation{Institute for Materials Science, University of Stuttgart, Pfaffenwaldring 55, 70569 Stuttgart, Germany}
\email{i.novikov@skoltech.ru}

\author{Blazej Grabowski}
\affiliation{Institute for Materials Science, University of Stuttgart, Pfaffenwaldring 55, 70569 Stuttgart, Germany}

\author{Fritz K\"{o}rmann}
\affiliation{Computational Materials Design, Max-Planck-Institut f\"ur Eisenforschung GmbH, D-40237 D\"usseldorf, Germany}
\affiliation{Materials Science and Engineering, Delft University of Technology, 2628 CD, Delft, The Netherlands}

\author{Alexander Shapeev}
\affiliation{Skolkovo Institute of Science and Technology, Skolkovo Innovation Center, Nobel St. 3, Moscow 143026, Russia}

\date{\today}

\begin{abstract}
We present the magnetic Moment Tensor Potentials (mMTPs), a class of machine-learning interatomic potentials, accurately reproducing both vibrational and magnetic degrees of freedom as provided, e.g., from first-principles calculations. The accuracy is achieved by a two-step minimization scheme that coarse-grains the atomic and the spin space. The performance of the mMTPs is demonstrated for the prototype magnetic system bcc iron, with applications to phonon calculations for different magnetic states, and molecular dynamics simulations with fluctuating magnetic moments. 

\end{abstract}

\maketitle

\section*{INTRODUCTION}
Magnetic contributions are essential for modelling magnetic materials as they critically affect phase stability \cite{hasegawa1983microscopic,herper1999ab,kormann2016influence}, vibrational properties \cite{ruban2012spin,ikeda2014phonon, kormann2012atomic}, interstitial energies \cite{dudarev2005magnetic}, local \cite{boukhvalov2007magnetism, gorbatov2011vacancy} and extended defects \cite{lu2017stacking, bienvenu2020impact}, and kinetics \cite{hegde2020atomic, schneider2020atomic}. Taking the magnetic degrees of freedom properly into account is a prerequisite for computationally-aided design and development of a large number of technologically relevant materials, ranging from various steels for construction and safety applications \cite{hasegawa1983microscopic,herper1999ab,kormann2016influence,ruban2012spin,ikeda2014phonon,kormann2012atomic,boukhvalov2007magnetism, gorbatov2011vacancy,lu2017stacking, bienvenu2020impact,hegde2020atomic, schneider2020atomic} to hard magnets for applications in electrical transportation and renewable energy technologies \cite{sozen2019ab, matsumoto2020calculating}.

One of the most popular computational methods, capable of capturing magnetism, are first-principles calculations realized by density functional theory (DFT). DFT calculations are, however, computationally expensive and limited to small system sizes and to a small number of magnetic configurations. DFT calculations that sample the magnetic degree of freedom explicitly, as needed for, e.g., lattice vibrations or vacancy formation energies in magnetically excited states, are therefore available only for very few selected cases. 

Recent progress in machine-learning potentials has significantly accelerated accurate simulations of materials and molecules \cite{behler2017-review,deringer2019-review,schutt2017-schnet,lubbers2018-DNN,vandermause2020-bayesian-on-the-fly,drautz2019-ace,smith2017-ANI,jinnouchi2019-on-the-fly,cusentino2020-snap-multielement,chmiela2017-finite-pes,pun2019-mishin-ann}. 
Such potentials express the interatomic energy as a function of atomic positions alone.
Ignoring the electronic degrees of freedom, yet assuming a very flexible functional form for the interatomic energies, machine-learning potentials feature near-quantum mechanical accuracy at a computational efficiency of the order of classical interatomic potentials \cite{zuo2020-benchmark}.
However, by ignoring the electronic degrees of freedom such potentials cannot distinguish different magnetic states, simply because different magnetic states feature different energies and the functional form of machine-learning potentials prohibits to capture such a magnetically-induced energy variation. In this paper we introduce a strategy to overcome this fundamental shortcoming.

\section*{Results}

\subsection*{Magnetic Moment Tensor Potential} \label{sec:mMTP}

The starting point is a given set of energies, $E^\dft(\bm{R},{S})$, which include the magnetic degree of freedom, e.g., computed via DFT, and where $\bm{R}$ and $S$ denote a set of atomic coordinates and corresponding atomic spins. There are various ways to compute these energies from DFT, e.g., via fully relaxing the spin degree of freedom or, if one is interested in a broader sampling of $E^\dft(\bm{R},{S})$, via constrained spin calculations \cite{dederichs1984,stocks1998towards,singer2005constrained,kaduk2012constrained,ma2015constrained}. We on purpose do not discuss in this work the different approaches available and their corresponding challenges to compute $E^\dft(\bm{R},{S})$ since the main focus here is on an efficient parametrization for a given $E^\dft(\bm{R},{S})$. We utilize standard spin-polarized DFT calculations where the local atomic moments are differently initialized while their longitudinal component is fully relaxed. The different magnetic configurations sampled are discussed further below. We note, however, that the proposed machine-learning potential can be straightforwardly applied with, e.g., constrained spin calculations.

The heart of the proposed approach is to
approximate the energy $E^\dft(\bm{R},{S})$ with 
Moment Tensor Potentials (MTPs) \cite{shapeev2016-mtp,gubaev2019-alloys} the idea of which is to expand the energy locally as a polynomial of its degrees of freedom, corrected in order to allow for a finite cutoff of the potential.
We note that there are other functional forms allowing for approximation of $E^\dft$ as a function of enriched degrees of freedom \cite{grisafi2018-GAP-tensorial,drautz2020-ACE-tensorial,nikolov2021-SNAP-Heisenberg}.
A similar functional form as utilized in MTPs has been recently employed within the atomic cluster expansion (ACE) \cite{bachmayr2019-ACE-completeness}. Both approaches feature a complete basis of invariant polynomials that differ only in the representation of the angular terms; MTP uses tensors while ACE uses spherical harmonics.

In our approach the total interaction energy is partitioned into contributions of individual local atomic environments:
\begin{equation} \label{eq:e-mtp}
E^{\mmtp} = \sum_{i=1}^N V(\mathfrak{n}_i),
\end{equation}
where $\mathfrak{n}_i$ is the neighborhood of the $i$'th atom and $N$ is the number of atoms in the atomic configuration.
In the present paper the degrees of freedom are atomic positions $\bm{R} = \{ \bm{r}_i, ~i=1,\ldots,N \}$ and spins $S = \{s_i, ~i = 1,\ldots,N \}$ as opposed to the originally developed MTPs \cite{shapeev2016-mtp,gubaev2019-alloys} in which the potential energy depends only on atomic positions. The atomic neighborhood of the $i$'th atom, $\mathfrak{n}_i$, is hence described by the relative interatomic positions $\bm{r}_{ij} = \bm{r}_j - \bm{r}_i$, the spin of the central atom, $s_i$, and the spins of the neighboring atoms $s_j$, formally
\[
{\rm \mathfrak{n}_i}=\{({\bm r}_{ij},s_i,s_j) : j=1,\ldots,N^i_{\rm nb}\},
\]
where $N^i_{\rm nb}$ is the number of neighbors of the $i$'th atom. 

The expansion of the function $V$ is:
\[
V(\mathfrak{n}_i) = \sum \limits_{\alpha} \xi_{\alpha} B_{\alpha}(\mathfrak{n}_i),
\]
where ${\bm \xi} = \{\xi_{\alpha} \}$ are the ``linear'' parameters to be optimized.
The function $V$ is assumed to be an arbitrary polynomial of the corresponding degrees of freedom, modified so that instead of the polynomial growth the potential $V$ vanishes beyond some cutoff distance.
The potential is expanded via basis functions $B_{\alpha}$
defined through the so-called moment tensor descriptors 
\begin{equation} \label{eq:descriptors}
M_{\mu,\nu}(\mathfrak{n}_i)=\sum_{j=1}^{N^i_{\rm nb}} f_{\mu}(|\bm{r}_{ij}|,s_i,s_j) \underbrace {\bm{r}_{ij}\otimes...\otimes \bm{r}_{ij}}_\text{$\nu$ times},
\end{equation}
where $``\otimes"$ is the outer product of vectors, and, thus, the angular part $\bm{r}_{ij}\otimes...\otimes \bm{r}_{ij}$ is a tensor of $\nu$'th rank.
The function $f_{\mu}(|\bm{r}_{ij}|,s_i,s_j)$ is a polynomial of $|\bm{r}_{ij}|$, $s_i$ and $s_j$, modified for a finite cutoff radius.
It has the form:
\begin{equation}\label{eq:f_mu}
f_{\mu}(|\bm{r}_{ij}|,s_i,s_j) = \sum_{\zeta=1}^{N_{\varphi}} \sum_{\gamma=1}^{N_{\psi}} \sum_{\beta=1}^{N_{\psi}} c^{\beta, \gamma, \zeta}_{\mu} \psi_{\beta}(s_i) \psi_{\gamma}(s_j) 
\varphi_\zeta (|\bm{r}_{ij}|) (r_{\rm cut} - |\bm{r}_{ij}|)^2,
\end{equation}
where ${\bm{c}}=\{c^{\beta, \gamma, \zeta}_{\mu} \}$ are the ``radial'' parameters to be optimized,  $N_{\varphi}$ is the number of polynomial basis functions $\varphi_{\zeta} (|\bm{r}_{ij}|)$ on the interval $[r_{\rm min}, r_{\rm cut}]$, where $r_{\rm min}$ is the minimal distance between atoms and $r_{\rm cut}$ is the cutoff radius beyond which atoms do not interact.
The term $(r_{\rm cut} - |\bm{r}_{ij}|)^2$ ensures a smooth vanishing of the potential for $|\bm{r}_{ij}| > r_{\rm cut}$.
The other functions, $\psi_{\beta}(s_i)$ and $\psi_{\gamma}(s_j)$, are the polynomial basis functions of the local spins of the central and neighboring atoms, respectively. The number of these spin basis functions is $N_{\psi}$.
They are defined on the interval $[s_{\rm min}, s_{\rm max}]$, where the values $s_{\rm min}$ and $s_{\rm max}$ are the minimal and maximal local magnetic moments in the system being investigated.

The mMTP basis functions $B_{\alpha}$ are defined as all possible contractions of $M_{\mu,\nu}(\mathfrak{n}_i)$ to a scalar, e.g.,
\[
M_{1,0}(\mathfrak{n}_i), ~M_{0,1}(\mathfrak{n}_i) \cdot M_{1,1}(\mathfrak{n}_i),  ~M_{3,2}(\mathfrak{n}_i):M_{1,2}(\mathfrak{n}_i), ~\ldots\,,
\]
where $``\cdot"$ is the dot product of two vectors, and $``:"$ is the Frobenius product of two matrices. In principle, an infinite number of such mMTP basis functions could be constructed. In order to choose which basis functions to include in practice in the mMTP, we introduce the so-called level of each descriptor, ${\rm lev} M_{\mu,\nu} = 2 + 4 \mu + \nu$, choose a certain $\rm {lev}_{\rm{max}}$, and include in the mMTP each basis function with ${\rm lev} B_{\alpha} \leq \rm {lev}_{\rm{max}}$ (see Ref. \cite{mlip-main-paper2020} for details).
Thus, the number of the ``linear" parameters $\bm \xi$ depends on $\rm {lev}_{\rm{max}}$, which also determines the number of radial functions, $N_{\mu}$.
The number of the ``radial" parameters $\bm c$ is equal to $N_{\mu} N_{\varphi} N_{\psi}^2$. We denote all free parameters of an mMTP collectively by $\bm \theta = \{\bm \xi, \bm c \}$, and the total interaction energy by $E^{\mmtp} = E^{\mmtp}(\bm \theta; \bm{R}, S)$.

We note that the mMTP formalism contains the Heisenberg model as a special, limiting case. In particular, first-degree polynomials have to be utilized for $\psi_\beta(s) = \psi_\gamma(s) = s$ in Eq.~\eqref{eq:f_mu}, and $\varphi_\zeta$ needs to ``encompass'' (i.e., be nonzero at) the nearest neighbors only.
	Such a choice of terms in the expansion Eq.~\eqref{eq:f_mu} also leads to a model similar to the one proposed in Ref.~\cite{nikolov2021-SNAP-Heisenberg}, except that in the latter case the full vectorial spins were considered.
	Moreover, the biquadratic terms, $(s_i s_j)^2$ \cite{rosengaard1997-biquad-Heisenberg,szilva2013-biquad-Heisenberg}, adopted by Ref.~\cite{nikolov2021-SNAP-Heisenberg}, arise naturally when $M_{\mu,0}$ is constructed with such choices of $\psi_\beta$ and $\varphi_\zeta$ and gets multiplied by itself.
	Then the radial parameters $c^{\beta, \gamma, \zeta}_{\mu}$ correspond to the coupling constants as obtained from DFT data.

The free parameters $\bm \theta$ in our approach are found by fitting $E^{\mmtp}$ to DFT data. We consider a training set including $K$ magnetic configurations $(\bm{R}^{(k)},{S}^{(k)})$ with known DFT energies $E^{{\dft}}$, DFT forces $\bm f^{{\dft}}_{i}$ on every atom $i$, and a $3\times 3$ tensor of DFT stresses $\sigma^{{\dft}}$ and minimize the objective function:
\begin{align*}
\sum _{k=1}^K \Biggl[ w_{\rm e} \Biggl|E^\mmtp\biggl(\bm \theta;\bm{R}^{(k)},{S}^{(k)}\biggr) - E^\dft\biggl(\bm{R}^{(k)},{S}^{(k)}\biggr)\Biggr|^2
\\
	\ + w_{\rm f} \sum_i \Biggl|\bm{f}^\mmtp_{i}\biggl(\bm \theta;\bm{R}^{(k)},{S}^{(k)}\biggr) - \bm{f}^\dft_{i}\biggl(\bm{R}^{(k)},{S}^{(k)}\biggr)\Biggr|^2
\\
\ +w_{\rm s} \Biggl|\sigma^\mmtp\biggl(\bm \theta;\bm{R}^{(k)},{S}^{(k)}\biggr)-\sigma^\dft\biggl(\bm{R}^{(k)},{S}^{(k)}\biggr)\Biggr|^2
\Biggr]
\end{align*}
where $|\cdot|$ is the length of a vector or the Frobenius norm of a matrix. The optimization of the parameters is carried out using an iterative quasi-Newton optimization method, specifically, the Broyden-Fletcher-Goldfarb-Shanno algorithm (BFGS) starting with a random initial guess. As opposed to mMTP, the energy of the non-magnetic MTP, proposed in our earlier works, does not depend on spins, i.e. $E^{\mtp} = E^{\mtp}(\bm \theta;\bm{R})$, and the functions $f_{\mu}(|\bm{r}_{ij}|)$ do not include spins.

\subsection*{Convergence of magnetic MTP}

\begin{figure}[tb!] \begin{center}
\includegraphics[width=0.45\linewidth]{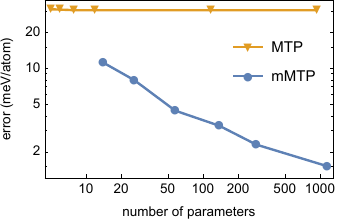}
\caption{\label{fig:error-convergence}
		Convergence of the magnetic potential, mMTP, with respect to DFT energies and lack of convergence for the non-magnetic MTP. The graph indicates that most variation of the energy on the training set is in the magnetic degrees of freedom that can be captured only by the magnetic potential.
	}
\end{center} \end{figure}

We first analyze the convergence behavior of the magnetic and non-magnetic MTP toward DFT energies as the number of parameters is increased.
The convergence was measured on a hold-out set of about 1000 configurations not participating in the fitting of the potentials.
Figure~\ref{fig:error-convergence} shows that the mMTP exhibits a steady convergence, while the non-magnetic MTP does not.
This reiterates our original motivation: the space of atomic positions $(\bm{R})$ is not the right one for approximating the quantum-mechanical energy, but enriched with spins, $(\bm{R},{S})$, this becomes a suitable space for that purpose.

Based on the convergence tests, we have chosen a well converged $\rm {lev}_{\rm{max}} = 24$ for the subsequent tests.
For both MTP and mMTP we took $N_{\varphi} = 12$ polynomial functions of the atomic positions with $r_{\rm min} = 2 ~\angstrom$, $r_{\rm cut} = 5.5 ~\angstrom$.
For the mMTP we took $N_{\psi} = 2$ polynomial functions of the local magnetic moments with $s_{\rm min} = -3.5 ~\mu_B$ and $s_{\rm max} = 3.5 ~\mu_B$.
The total number of MTP parameters was 937 while that of mMTP was 1153.
The weights in the objective function were $w_{\rm e} = 1$, $w_{\rm f} = 0.01$, and $w_{\rm s} = 0.001$.

For each model we fitted five potentials and selected the best (with the least training error).
The validation root-mean-square errors are shown in Table~\ref{tabl:ave-training-errors}.
We can see that adding local magnetic moments to the potential as additional degrees of freedom does not significantly increase the number of parameters, but greatly improves the accuracy of training.

\subsection*{Phonon spectra prediction}
\begin{figure*}[htb!] \begin{center}
\includegraphics[width=0.9\linewidth]{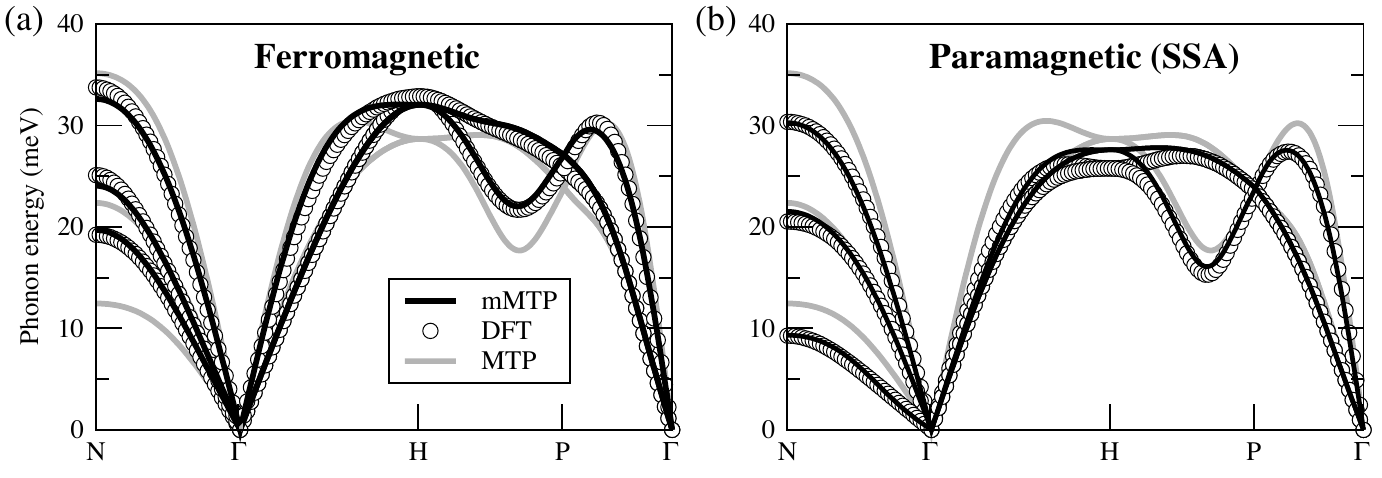}
\caption{\label{fig:phonon-spectra}
Computed phonon spectra for the (a) ferromagnetic and (b) paramagnetic state (modelled using the SSA approach) with DFT, non-magnetic MTP, and magnetic MTP.}
\end{center} \end{figure*}

We next evaluate the performance of the best optimized MTP and mMTP potentials to predict phonon spectra of different magnetic states. We consider two extreme scenarios representing the limits of magnetic configurations, namely the ferromagnetic state, in which all spins are aligned parallel and a paramagnetic state, treated in the adiabatic limit of fast fluctuating spins. Since the phonon energies were derived from small perturbations (utilizing the small displacement method), this test is a very sensitive measure to detect how well even very small variations in interatomic forces can be captured. The results for the ferromagnetic case for both potentials are shown in Figure~\ref{fig:phonon-spectra}(a) in comparison with the data directly obtained from DFT. The agreement between the mMTP and the DFT data is excellent whereas the non-magnetic MTP shows significant deviations, in particular around the N-point. The deviations for the non-magnetic MTP are a direct consequence of the training database which also includes magnetically disordered configurations responsible for pronounced phonon softening as discussed in the following. 

\begin{table*}
	\begin{center}
		\begin{tabular}{|r|c|c|c|c|c|} \hline 
			\multicolumn{1}{|c|}{\multirow{2}{*}{model}} & \multicolumn{1}{|c|}{\multirow{2}{*}{level}} & number of  & energy error & force error & stress error \\
			& & parameters & meV/atom & meV/\angstrom\ (\%) & GPa ($\%$) \\ \hline
			MTP  & 24 & 937 & 30.4 & 195 (27.3 \%) & 0.542 (13.3 \%)  \\ \hline
			mMTP & 24 & 1153 & 1.5 & 64 (9.0 \%) & 0.087 (2.2 \%) \\ \hline
		\end{tabular}
		\caption{\label{tabl:ave-training-errors}
				The best non-magnetic and magnetic MTPs. Despite the fact that the magnetic MTP has only a small increase in the number of parameters, its accuracy is much higher than that of the non-magnetic MTP.
			}
	\end{center}
\end{table*}

To compute the phonon spectra in the paramagnetic regime we utilize the spin-space averaging (SSA) method \cite{kormann2012atomic}. In this approach effective interatomic forces can be defined by averaging over various disordered magnetic configurations weighted by a Boltzman distribution. For the actual averaging we utilized the crystal symmetries as proposed in Refs.~\cite{kormann2012atomic,ikeda2014phonon} and performed the SSA using a single random magnetic configuration for which each atom is displaced in each cartesian direction. This provides a large number of locally inequivalent magnetic configurations (i.e., $54\cdot 3=162$ configurations for the employed supercell).
	This procedure was shown to be robust with respect to the actually chosen random magnetic configuration as discussed in Ref.~\cite{kormann2012atomic}.

The resulting DFT based phonon spectrum shown in Figure~\ref{fig:phonon-spectra}(b) features a pronounced softening at the N-point \cite{kormann2012atomic}. This softening is related to the decrease of the elastic constants and constitutes an important precursor of the structural transformation in iron. The non-magnetic MTP cannot distinguish the underlying atomic forces in these different magnetic states from the ferromagnetic forces. This is the reason why the MTP phonon spectrum for the paramagnetic state shown in Figure~\ref{fig:phonon-spectra}(b) is exactly the same as the one in Figure~\ref{fig:phonon-spectra}(a) for the ferromagnetic state. The non-magnetic MTP spectra fall in-between the ferromagnetic and paramagnetic solutions and hence do not quantitatively reproduce the DFT data in either regime. In contrast, applying the SSA approach with the mMTP reveals an excellent agreement with the DFT data, reproducing quantitatively important characteristics such as, e.g., the decrease of the phonon energies near the N-point and along the H-P path. 

\subsection*{Disordered-local-moment molecular dynamics simulations}

To evaluate the performance of the mMTP at finite temperatures and larger atomic displacements, we have performed molecular dynamics (MD) simulations. The temperature was set to 800~K and the lattice constant to 2.9 \angstrom. To sample not only the vibrational degrees of freedom but the spin space and in particular the coupling between vibrations and spins, we have performed disordered-local-moment MD (DLM-MD) simulations \cite{alling2016strong}. Further, in order to explicitly validate the mMTP against DFT, we have utilized the concept of thermodynamic integration, similarly as used in the TU-TILD+MTP method previously \cite{grabowski2019-hea}. Specifically, we have introduced a linear coupling between DFT and mMTP forces,
\begin{align}
\label{eq:TI-forces}
    F_\lambda = \lambda F^\textrm{DFT} + (1-\lambda) F^\textrm{mMTP},
\end{align}
with the coupling constant $\lambda$ and DFT and mMTP forces $F^\textrm{DFT}$ and $F^\textrm{mMTP}$. The coupled forces $F_\lambda$ were used for evolving the DLM-MD trajectories. The mMTP in this test was fitted to 'pure' DFT calculations (i.e., nominally corresponding to $\lambda=1$) and tested independently for a new set of calculations at $\lambda=0,0.5,1$. To render the DFT calculations feasible we employed a 16-atom supercell for the DLM-TI calculations; cross-checks for a 54-atom supercell showed similar results. Further details are given in the Methods section.

\begin{figure}[tb!] \begin{center}
\includegraphics[width=1\linewidth]{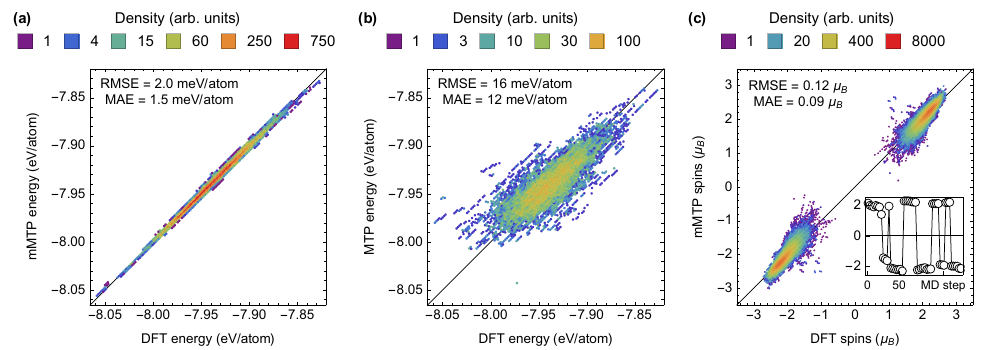}
\caption{\label{fig:correlations} Energy correlation (a) between magnetic MTP and DFT, (b) between non-magnetic MTP and DFT, and (c) spin correlation between magnetic MTP and DFT from DLM-TI calculations at 800 K.}
\end{center} \end{figure}

Figure~\ref{fig:correlations} highlights the excellent performance of the mMTP. In the left panel we observe that the mMTP energies fall almost on top of the DFT energies; the root-mean-square error (RMSE) is only 2.0 meV/atom---of the same order as obtained previously for non-magnetic systems \cite{grabowski2019-hea}. The middle panel clarifies that the best possible non-magnetic MTP is almost an order of magnitude away in terms of energy accuracy, with an RMSE of 16 meV/atom. The right panel of Figure~\ref{fig:correlations} shows the spin correlation between mMTP and DFT, which of course only the magnetic version of the MTP is capable to reproduce. We observe an RMSE of 0.12 $\mu_B$, which is about 5\% of the magnitude of the absolute spin.

We stress that Figure~\ref{fig:correlations} includes values for all the investigated coupling constants $\lambda=0.0$, $0.5$, $1.0$. Looking at each $\lambda$ value separately, the correlations are in fact very similar. This means that there is no difference in the correlation, if we use pure DFT forces (cf. Eq.~\eqref{eq:TI-forces}), pure mMTP forces, or DFT-mMTP coupled forces to evolve the MD. This hence allows one to perform a full thermodynamic integration from the mMTP to DFT and compute the respective free energy difference, which is however beyond the scope of the present work.

\section*{DISCUSSION}

We have developed the mMTPs, a class of magnetic machine-learning interatomic potentials capable of simultaneously and accurately approximating spin and atomic degrees of freedom.
This has been achieved by utilizing a two-step minimization scheme for the spin and atomic configurational space. Applying the mMTP to DFT-derived data for the prototypical bcc iron system reveals that the mMTPs are capable to quantitatively approximate local magnetic moments, energies, and forces for various magnetic states (see the Supplementary Discussion for further tests). A number of applications such as the computation of phonon spectra in ferro- and paramagnetic states as well as molecular-dynamics simulations including spin-flips demonstrate that mMTPs provide near DFT-accuracy without significantly losing the computational efficiency of classical interatomic potentials. 

\section*{Methods}
\subsection*{Derivation of \lowercase{m}MTP}

Here we derive the form of the MTP as a function of relative atomic positions $\bm{r}_{ij}$ and vectorial magnetic moments, $\bm{s}_{i}$ and $\bm{s}_{j}$.
Following the logic of the original paper introducing the MTP \cite{shapeev2016-mtp}, an arbitrary polynomial of the positions and magnetic moments can be represented as all possible contractions of the following Moment Tensors,
\begin{equation}\label{eq:M-general}
M_{\zeta,\nu,\beta,\gamma,\xi,\eta} = \sum_{j=1}^{N^i_{\rm nb}} Q_{\zeta}(|\bm{r}_{ij}|) |s_i|^\beta |s_j|^\gamma
\,
\big(\bm{r}_{ij}^{\otimes \nu}\big) \!\otimes\!
\big(\bm{s}_{i}^{\otimes \xi}\big) \!\otimes\!
\big(\bm{s}_{j}^{\otimes \eta}\big)
,
\end{equation}
where $Q_\zeta$ is the $\zeta$-th radial basis function and by definition
\[
\bm{v}^{\otimes n} = \underbrace{\bm{v}\otimes...\otimes \bm{v}}_\text{$n$ times}
\]
for an arbitrary vector $\bm{v}$.
We note that Eq.~\eqref{eq:M-general} directly corresponds to Ref.~\cite[Eq.\ 26]{drautz2020-ACE-tensorial} with spherical harmonics $\bm{Y}_l^m(\bm{v}/|v|)$ instead of tensors $\bm{v}^{\otimes n}$ ($|l| \leq m \leq n$).

We next assume scalar-valued spins, i.e., that it is sufficient to consider $\xi=\eta=0$ (and adsorb the sign of the spin into the radial part):
\begin{equation}\label{eq:M-scalar}
	M_{\zeta,\nu,\beta,\gamma} = \sum_{j=1}^{N^i_{\rm nb}} Q_{\zeta}(|\bm{r}_{ij}|) s_i^\beta s_j^\gamma
	\, \big(\bm{r}_{ij}^{\otimes \nu}\big)
	.
\end{equation}
We could choose to directly expand the energy over different contractions of tensors $M_{\zeta,\nu,\beta,\gamma}$, but instead we combine different products of radial and spin basis functions, $Q_{\zeta}(|\bm{r}_{ij}|) s_i^\beta s_j^\gamma$, into the functions
$f_\mu(\bm{r}_{ij},s_i,s_j)$ with coefficients $c^{\beta, \gamma, \zeta}_{\mu}$ that are found from data, as explained in the main text of the manuscript.
Note that in this work we use Chebyshev polynomials $\psi_\beta(s)$ instead of monomials $s^\beta$.

\subsection*{DFT details} \label{sec:TrainingSet}

All DFT calculations were performed with \textsc{vasp} \cite{VASP1, VASP2,VASP3,VASP4} utilizing the projector augmented wave (PAW) method \cite{blochl1994projector} and the generalized gradient approximation \cite{perdew1996generalized}. For the training set of the 54-atom supercell we considered 70 atomic configurations generated from an initial ferromagnetic MD at 1000 K. For each of these atomic configurations 200 different  arrangements of magnetic spins have  been initialized of which 67$\%$ converged under the high cutoff energy of 500~eV and $k$-point density of 11664 $k$-points$\times$atoms (6$\times$6$\times$6 grid) chosen in combination with a convergence criterion of $10^{-7}$ eV per supercell to ensure high-accurate DFT data. This resulted into in total 9351 calculations. To impose spin-inversion symmetry we added the same number of configurations to the training with reversed spin directions.  The DFT calculations were performed at a lattice parameter of 2.9 \angstrom\ corresponding to the experimental value near the Curie temperature. The phonon calculations have been performed utilizing the finite-displacement method with a displacement of 0.02~\AA{} and utilizing the same set of technical parameters.

\subsection*{Disordered-local moment thermodynamic integration from \lowercase{m}MTP to DFT}

To sample the paramagnetic state at finite temperatures within the framework of thermodynamic integration (TI), we have employed the disordered-local-moment (DLM) MD \cite{alling2016strong}. The local magnetic moments were flipped randomly every 10 fs ($=$\,10 MD steps) such that half of the moments was pointing up and the other half down. The timestep for the MD was set to 1 fs; small enough to sample well the time development of the magnetic moments also within the 10 fs time intervals. The temperature was controlled by the Nose thermostat \cite{Nose}. Usage of the Nose thermostat was critical; tests with the Langevin thermostat showed that it cannot stabilize the temperature well due to the additional impact of the spin flips on the energy of the system.

Spin-polarized DFT calculations in general and DLM calculations for Fe in particular are very prone to convergence problems, due to a flat energy landscape with many local minima as a function of spin state. Therefore, for the calculation of the DFT energy and forces during the MD, a very tight convergence criterion of $10^{-7}$ eV per supercell was set, in order to enforce sampling of the original DFT energy landscape that served as the input to the magnetic MTP fitting. To nevertheless allow for an efficient DFT MD, we have restricted the number of electronic iteration steps (typically to 40). Not fully converged DFT calculations were omitted from the comparison to the magnetic mMTP. Likewise DFT calculations featuring local moment flips with respect to the mMTP data were not considered in the comparison.

To increase the efficiency of the DFT DLM-MD simulations we found it necessary to turn off the wave function extrapolation (both linear and quadratic); the reason for this lying in the randomization of the spins along the MD trajectory. A further efficiency increase was achieved by equilibrating the MD at the temperature of interest by utilizing the efficient mMTP. In this way the part of the MD involving the expensive DFT calculations started directly on a well equilibrated trajectory.

The DLM-TI was performed at a temperature of 800 K and at a lattice constant of 2.9 \AA{}. A supercell of 2$\times$2$\times$2 (in terms of the conventional bcc unit cell) with 16 atoms was utilized. A dense $k$-point sampling of 6$\times$6$\times$6 corresponding to 3,456 $k$-points$\times$atom, a plane wave cutoff of 500 eV, and Fermi-Dirac smearing were used for the DFT calculations. For the mMTP calculations, initial magnetic moments were set according to the DFT moments. Then, for every mMTP energy and force calculation, the magnetic moments were fully relaxed based on the mMTP energetics. Coupling constants of $\lambda=0.0, 0.5, 1.0$ were utilized. At each coupling constant two different random seeds were used to generate distinct trajectories. In total more than 22,000 of MD steps (22 ps) were conducted to generate statistically highly reliable correlation plots as shown in Figure~\ref{fig:correlations} of the main text. Test calculations for a larger 3$\times$3$\times$3 supercell with 54 atoms turned out to be computationally highly demanding due to the strict DFT convergence parameters. Corresponding results indicate however a similar performance of the mMTP also for the larger supercell.

\section*{DATA AVAILABILITY}

The datasets generated during and/or analyzed during the current study are available from the corresponding author on reasonable request.

\section*{ACKNOWLEDGEMENTS}

We acknowledge support from the collaborative DFG-RFBR Grant (Grants No. DFG KO 5080/3-1, DFG GR 3716/6-1, and RFBR 20-53-12012). B.G. acknowledges the support by the Stuttgart Center for Simulation Science (SimTech) and funding from the European Research Council (ERC) under the European Union’s Horizon 2020 research and innovation programme (grant agreement No 865855).

\section*{AUTHOR CONTRIBUTIONS}

    F.K. and A.S. conceived the project.
    I.N. and A.S. developed the magnetic Moment Tensor Potential (mMTP).
    I.N. fitted mMTP, investigated prediction errors and convergence of mMTP.
    F.K. prepared DFT datasets, calculated and compared the phonon spectra obtained with the magnetic MTP and DFT.
    B.G. conducted the disordered-local-moment molecular dynamics simulations and investigated the correlation of magnetic MTP and DFT energies and spins.
    All the authors participated in analyzing the results and writing the manuscript.
    
\section*{COMPETING INTERESTS}

The authors declare no competing interests.


%

\newpage

\section*{Supplementary Discussion}

\newcommand{\figurereflabel}{Supplementary Figure}

In the article we have illustrated the use of the magnetic moment tensor potential (mMTP) as an accurate approximant to DFT in phonon and molecular dynamics simulations of bcc ferromagnetic and paramagnetic iron. Here we provide the results of additional tests: predicting with mMTP the vacancy formation energy (VFE) for bcc iron in the ferromagnetic state and the energy/volume curves for bcc, fcc, and hcp iron in the ferromagnetic and antiferromagnetic states.

To compute the VFE in bcc iron in the ferromagnetic state we created a training set containing 71 configurations of 54 atoms (including the equilibrium one) and 47 configurations of 53 atoms (i.e., the configurations with the vacancy; the equilibrium one was also included). The training set was calculated with VASP \cite{VASP1,VASP2,VASP3,VASP4} using the settings and parameters described in the main text. We fitted mMTP with 1153 parameters on this training set. We have obtained a VFE of 2.28 eV with VASP and 2.19 eV with mMTP.

In order to fit mMTP for calculating energy-volume curves we used the following number of configurations in the ferromagnetic state: 24 bcc configurations (2-atomic unit cells), 23 fcc configurations (1-atomic unit cells), and 12 hcp configurations (2-atomic unit cells).
The dataset also contained antiferromagnetic configurations including 24 bcc configurations, 23 fcc configurations (cubic 4-atomic unit cells), and 12 hcp configurations. A high energy cutoff (500 eV) and $k$-point densities (always $>10,000$ $kp\cdot$atom) have been chosen.
We fitted mMTP with 172 parameters and calculated the energy volume/curves and magnetic moments in the equilibrium state using mMTP and DFT. They are shown in Supplementary Figures~1-2. Both energy-volume curves and magnetic moments computed with the fitted mMTP are close to the ones computed with DFT for bcc, fcc, and hcp iron in the corresponding ferromagnetic, antiferromagnetic, and nonmagnetic states.

\newpage

\begin{figure*}[htb!] \begin{center}
\includegraphics[width=6in, height=5in]{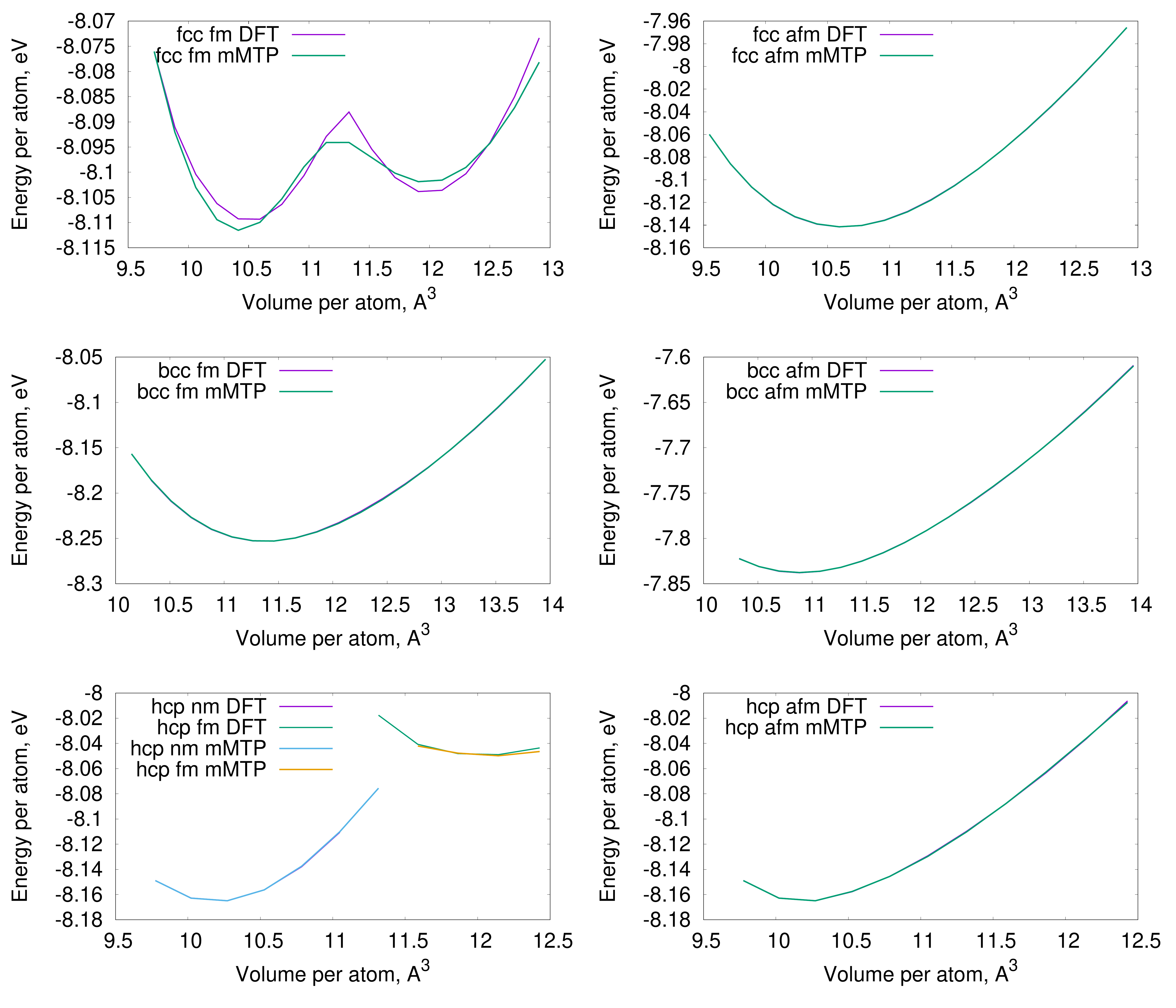}
\caption*{\figurereflabel{ 1:} Energy-volume curves computed with DFT and mMTP for bcc, fcc, and hcp iron in the ferromagnetic (fm), antiferromagnetic (afm) and nonmagnetic (nm) states. mMTP reproduces well all the corresponding magnetic states, including the low-spin to high-spin transition in fcc iron and the nonmagnetic to ferromagnetic transition in hcp iron.}
\end{center} \end{figure*}

\newpage

\begin{figure*}[htb!] \begin{center}
\includegraphics[width=6in, height=5in]{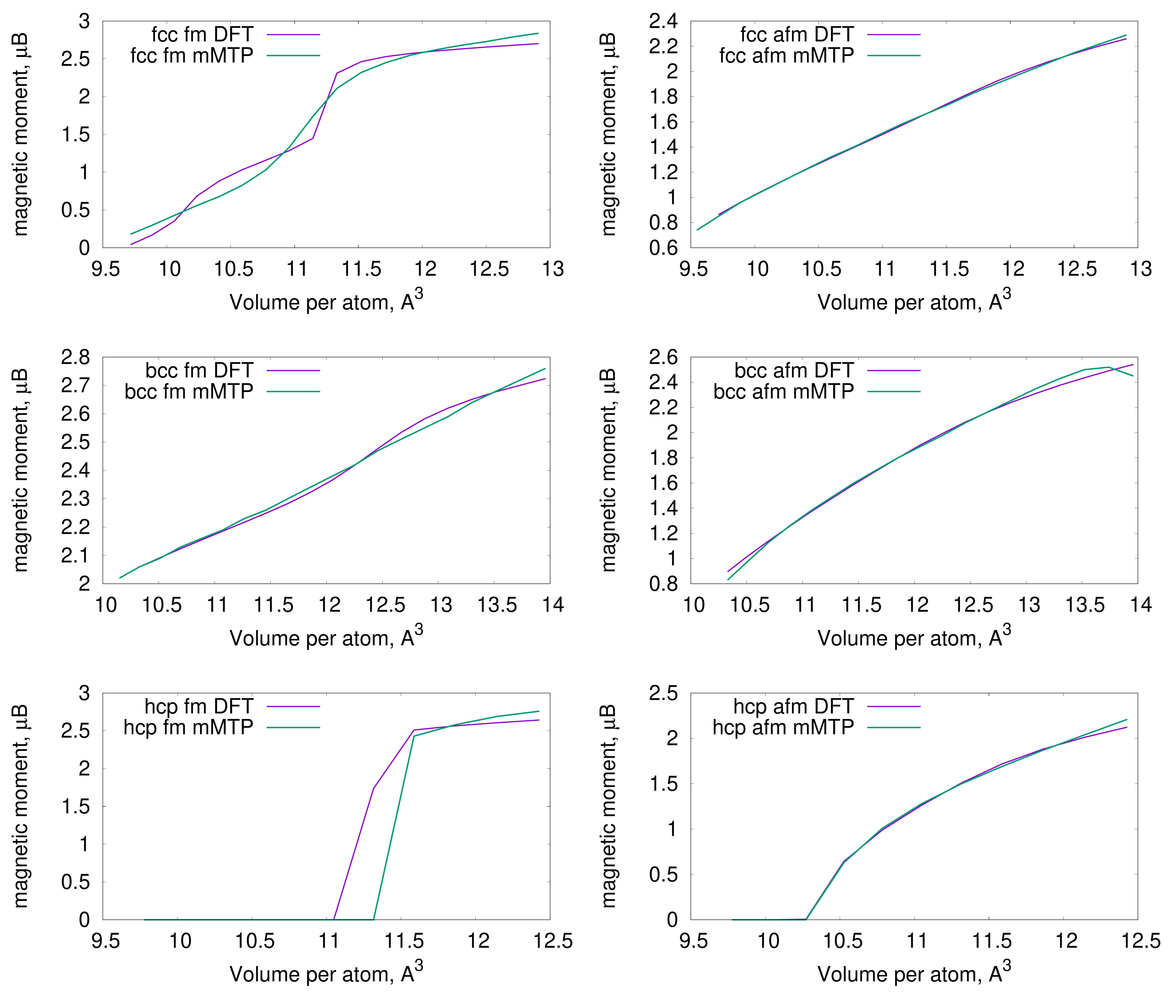}
\caption*{\figurereflabel{ 2:} Dependence of absolute equilibrium magnetic moments computed with DFT and mMTP on volume per atom for bcc, fcc, and hcp iron in the ferromagnetic and antiferromagnetic states.}
\end{center} \end{figure*}

\end{document}